\documentclass{article}
\newcommand{\ba}{\begin{eqnarray}}
\newcommand{\ea}{\end{eqnarray}}
\begin{document}
\begin{center}
{\bf Quantum Information as Reduced Classical Information}
\vspace*{0.5cm}\\ B.A.Nikolov
\\ {\it Department of Theoretical and Applied Physics, \\
Shoumen University, 9712 Shoumen, Bulgaria\\ E-mail:}
b.nikolov@shu-bg.net
\end{center}
\vspace*{0.5cm}

\begin{abstract}
It is shown on a simple classical model of a quantum particle at
rest that information contained into the quantum state (quantum
information) can be obtained by integrating the corresponding
probability distribution on phase space, i.e. by reduction of the
information contained into the  classical state.
\end{abstract}

1.A physical theory must give specification of three items: a)
state space, b) observables, and  c)the value of an observable in
a given state.For example in the classical model state space is
the phase space of the system,observables are smooth real
functions on phase space,and the value of the observable $A$ in a
state $(q,p)$ is $A(q,p)$.In general however the phase space point
is not known and the information about the real state (classical
information)is given by a phase space distribution $\rho(q,p)$.In
this case one can predict only the mean value (expectation)of the
observable \ba \langle A \rangle= \int A(q,p)\rho(q,p)dqdp \ea
Thus it is reasonable under a {\it(classical)} state to understand
not a point in phase space but rather a probability distribution
on it.Then instead about the value of an observable in some state
one should speak about the {\it mean} value of the observable in
that state.

In the quantum model the state space is a Hilbert space,the
information about the state ({\it quantum state})is represented by
a density operator ,i.e a positive operator $\rho$ satisfying $ tr
\rho=1 $,observables are Hermitian operators, and the mean value
of the observable $A$ in state $\rho$ is given by the Born rule
\ba \langle A\rangle =tr A \rho \ea .

Our aim in the present note is ,using  the  simple model of a
quantum particle given in a previous work \cite {Nik-03}, to show
that ,in fact, the quantum state can be obtained by integrating of
the classical one .

2.As a model of a quantum particle at rest we consider a $d =2s+1$
-dimensional oscillator ,where $s$ is the spin of the particle.The
mass and the frequency of the oscillator are taken equal to unit.
The coordinates $q_{n},(n=1,2,\dots,d)$ of the oscillator describe
the microscopic (unobservable) deviation from the rest position.If
$p_{n}$ are the corresponding momenta,and energy is assumed to be
$E=\hbar\omega/2= \hbar/2$ then the phase space trajectory is
given by the equation
 \ba \sum_{n=1}^{d}(p_{n}^{2}+q_{n}^{2})=\hbar \ea Introducing new
 coordinates and momenta , viz. $
 x_{n}=\hbar^{-1/2},y_{n}=\hbar^{-1/2},\psi_{n}=x_{n}+iy_{n}$ we
 can rewrite Eq.(1) in the form \ba \sum_{n}\psi_{n}^{*}\psi_{n}=
 1\ea Now using the same techniques as in Ref.1 we can
 show that arbitrary observable generating canonical
 transformation that preserve Eq.(4) is a hermitian form: \ba A(\psi^{*},\psi) =
 \hbar\sum_{n}A_{nm}\psi_{n}^{*}\psi_{n} \ea Here $A_{nm}$ is a
 Hermitian matrix completely determining the observable.

 3.Let $\rho(\psi^{*},\psi)$ is  the probability distribution of
 phase space coordinates, and \ba
 d\Omega_{\psi}=(d-1)!\delta(1-\psi^{*}\psi)d\psi^{*}d\psi \ea is
 the normalized measure over the unit sphere (4)
 \cite{91-Jon}.Then, applying Eq.(1) taken in the form \ba
 \langle A \rangle = \int A(\psi^{*},\psi)\rho(\psi^{*},\psi)d\Omega_{\psi}
 \ea we obtain \ba \langle A \rangle = \sum_{n,m}A_{nm} \rho_{mn}
 \ea  where \ba \rho_{mn} = \hbar \int \psi_{n}^{*}\psi_{m} \rho
 (\psi^{*},\psi)d\Omega_{\psi} \ea In fact we obtained the Born
 rule in its most general form.


\begin{thebibliography}{99}
\bibitem {Nik-03} B.A.Nikolov, "A Thermodynamic Approach to Quantum
Measurement and Quantum Probability", quant-ph 0303121.
\bibitem {91-Jon} K.R.W.Jones, Ann. Physics,v.207,p.140.
\end{thebibliography}
\end{document}